\documentclass[twocolumn,twoside,slac_two]{revtex4}
\usepackage{graphicx}
\usepackage{fancyhdr}
\setcitestyle{numbers}

\def \et {E_{T}}
\def \etx {E_{T,x}}
\def \ety {E_{T,y}}
\def  \met {\not\!\!\et }
\def  \metx {\not\!\!\etx }
\def  \mety {\not\!\!\ety }

\newcommand{\nuwt}	{$\nu$WT}
\newcommand{\nuwth}	{$\nu$WT$_h$}
\newcommand{\nuwtf}	{$\nu$WT$_f$}
\newcommand{\ltrk}	{$\ell+$track}

\begin{document}

\title{Top Mass Measurements with the D\O\ Detector}

%

\author{Daniel Boline}
\affiliation{Department of Physics, Boston University, Boston, MA 02215, USA \\ On behalf of the D\O\ Collaboration}

\begin{abstract}
I present recent results related to the measurement of the top quark mass, using $p\bar{p}$ collisions recorded with the D\O\ Detector at the Fermilab Tevatron Collider.  The results are: A direct measurement of the mass difference between top and anti-top quarks, Measurement of the top quark mass in the lepton+jets channel and in the dilepton channel.
\end{abstract}

\maketitle

\thispagestyle{fancy}


\section{Introduction}


The D\O\ detector, at the Tevatron, is a general purpose particle detector, composed of tracking, calorimeter, and muon systems.  The Tevatron collides opposing beams of protons and anti-protons, each with an energy of 0.98 TeV, the 1.96 TeV center of mass energy an energy makes the Tevatron the first, and still only, top quark factory.

Top quarks at the tevatron are produced both singly and in pairs.  However the single top production mode has both a smaller cross section, and more importantly a large irreducible background.  When produced in pairs, both top and anti-top decay into a $W$ boson and $b$ quark, the $W$ boson can then decay into a lepton and neutrino or a pair of quarks, and finally the $b$ quarks, and any light quarks ($u$,$d$,$s$) produced by the $W$ decay undergo particle showering and hadronization.



Since 2001 the Tevatron has produced $\sim 40000$ $t\bar{t}$ pairs, though only a fraction of those events are useful for the mass analyses.

\section{Lepton+jets Event Selection}

There are two lepton+jets channels used for the mass analysis, electron+jets (e+jets) and muon+jets ($\mu$+jets).  Events in both channels are required to have a single isolated lepton with $p_T>20\mbox{ GeV}$ and $|\eta|<1.1$ ($|\eta|<2$ for muons), to have four jets with $p_T>20\mbox{ GeV}$ and $|\eta|<2.5$, and to have large missing transverse energy $\met > 20\mbox{ GeV}$ ($\met > 25\mbox{ GeV}$ for muons).


The main backgrounds in the lepton+jets channels come from multijet QCD, and $W$+jets production.  The QCD signal is reduced by placing quality cuts on the leptons, and also by placing a cut on the azimuthal angle between the lepton and the $\met$.  The $W$+jets background is reduced by requiring that one jet be tagged using a neural network discriminant trained to distinguish b-jets from light-jets.

\section{Top Mass Difference Measurement}

At the Tevatron, we can measure the top mass to a high precision, using three different channels, all-jets, lepton+jets, and dileptons.  In all cases we assume that the mass of the top and anti-top are the same, such an equality is demanded by CPT (charge, parity, time) invariance~\cite{Cembranos:2006hj}.  However, we know that studies of C and P individually led to the discovery that neither symmetry is conserved by the weak interaction, and similarly a violation of CPT would require new physics to explain.

\subsection{Matrix Element Method}

The top mass measurement uses the matrix element method (ME).  The ME uses probability densities determined by integrating over the cross section for both the signal ($t\bar{t}$), and the main background ($W$+jets).  A likelihood is constructed having the form:

\begin{eqnarray}
 \mathcal{L}(m_{t},m_{\bar{t}},f_{t}) & & = \\
 \prod_{i=1}^{N_{evt}} & & \left[f_{t}P_{sig}(x_i;m_{t},m_{\bar{t}})+(1-f_{t})P_{bkg}(x_i)\right] \nonumber
\end{eqnarray}

Where $x_i$ represents the lepton and jet momenta measured in the detector and $f_t$ is the fraction of $t\bar{t}$ events in the sample.  The signal probability $P_{sig}$ has the form:

\begin{eqnarray}
 P_{sig}(x;m_t,m_{\bar{t}}) = Acc(x) \times \frac{1}{\sigma} & & \\
  \times \int d^n \sigma(y;m_t,m_{\bar{t}}) dq_1 dq_2 & \times & f(q_1) f(q_2) W(x,y) \nonumber
\end{eqnarray}

\begin{eqnarray}
d^n \sigma(y;m_t,m_{\bar{t}}) &=& d\Phi_6 \\
&\times & \frac{(2\pi)^4 |\mathcal{M}(q\bar{q}\rightarrow t\bar{t} \rightarrow y;m_t,m_{\bar{t}})|^2}{2q_1q_2s} \nonumber
\end{eqnarray}

Where $Acc(x)$ corrects for the acceptance of the event selection, $\sigma$ is the total cross section, y represents the momenta of the $t\bar{t}$ decay products, $|\mathcal{M}|^2$ is the matrix element of the process $q\bar{q}\rightarrow t\bar{t}$, $d\Phi_6$ is the 6-body phase space integral, $q_1$ and $q_2$ are the momenta of the incoming partons, and $W(x,y)$ is the transfer function which is the probability of parton level momenta $y$ producing the measured momenta $x$.  The background probability ($P_{bkg}$) has a similar form.

For convenience we transform the original variables ($m_t$,$m_{\bar{t}}$) are transformed into two new variables: $\Delta m_t = (m_t - m_{\bar{t}})$ and $m_{sum} = (m_t + m_{\bar{t}})/2$.  When quoting a final value for $\Delta m_t$, $m_{sum}$ is integrated out, with the final likelihood having the form:

\begin{equation}
 \mathcal{L}(\Delta m_t) = \int dm_{sum} \mathcal{L}(\Delta m_t , m_{sum})
\end{equation}

\subsection{Simulation of Signal and Background Samples}

\begin{figure}[h]
\includegraphics[width=80mm]{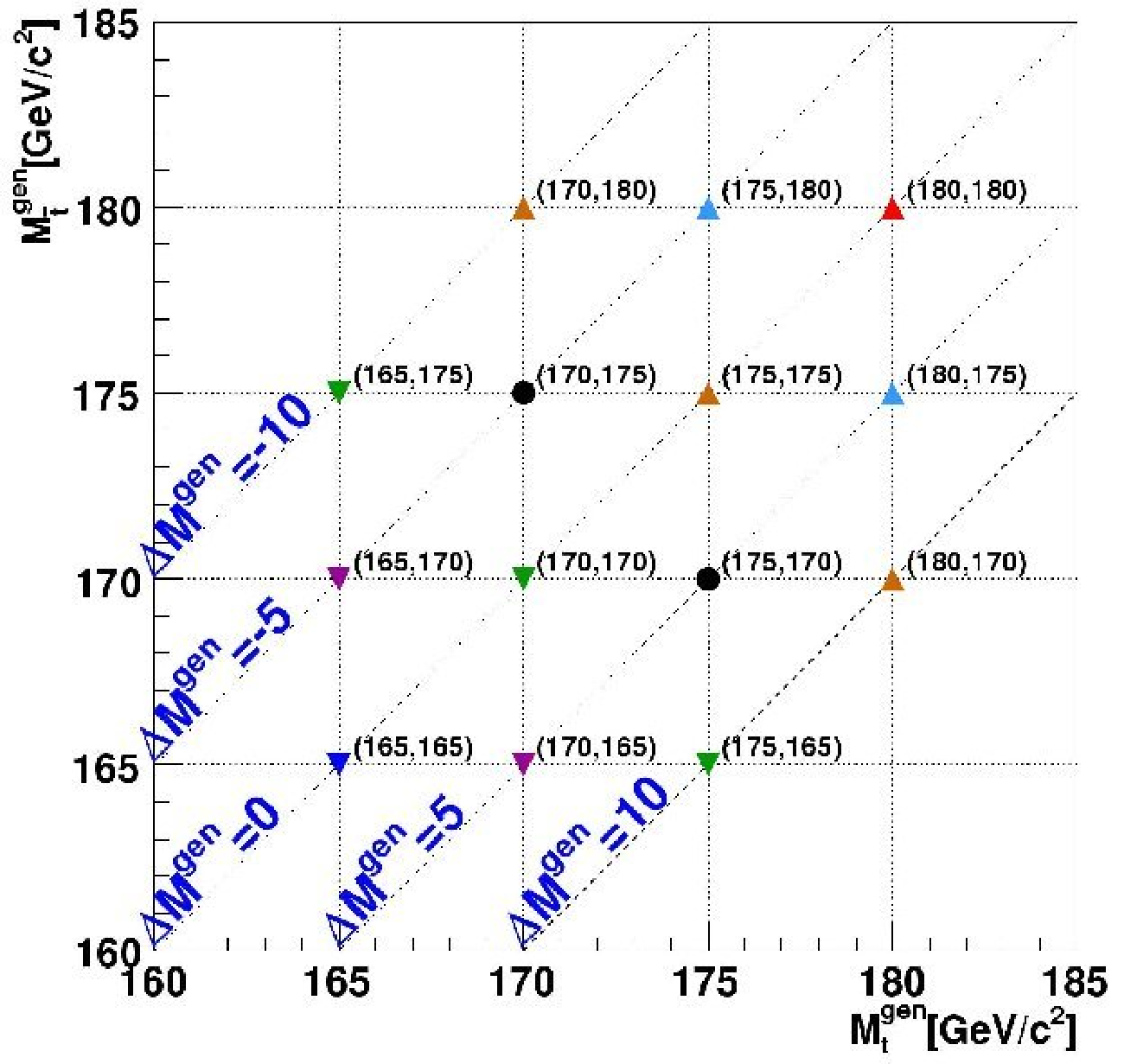}
\caption{\label{mc_gen_diff} $m_t$ and $m_{\bar{t}}$ values for simulated signal.}
\end{figure}

Signal samples for this analysis have been generated using the PYTHIA Monte Carlo event generator~\cite{Sjostrand:2000wi}.  Top and anti-top masses are given values of 165, 170, 175, and 180 GeV, with 14 samples in total (see Fig. \ref{mc_gen_diff}).

The $W$+jets background was generated using the leading order matrix element generator ALPGEN~\cite{Mangano:2002ea}, with PYTHIA used to shower the partons generated by ALPGEN.

\subsection{Performance on Simulated Data, Calibration of Method}

We have constructed ensembles of pseudoexperiments to test the performance of the method, and remove any bias on the measurement.  The pseudoexperiments are formed using simulated events, and contain signal and background events consistent with that expected in data.  The measurement procedure is performed for each pseudoexperiment in the same way as for data.

\begin{figure}[h]
\includegraphics[width=80mm]{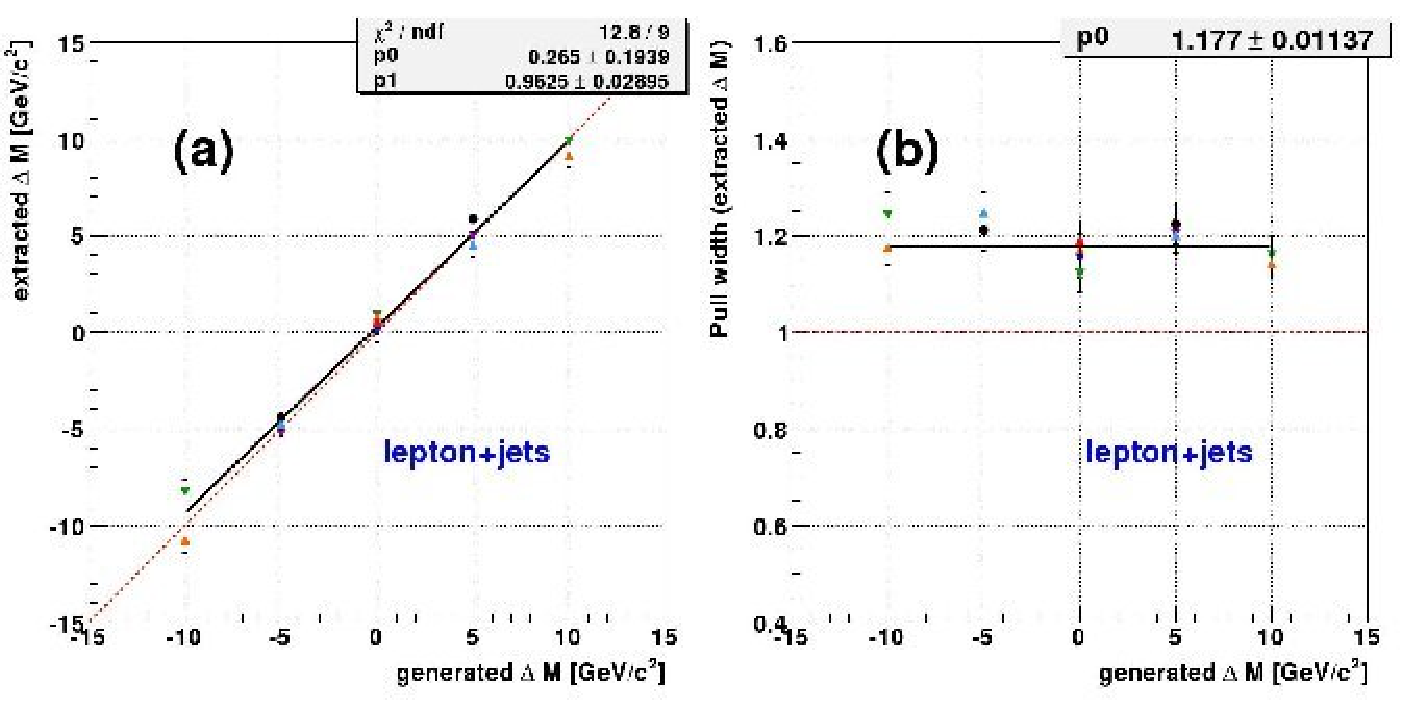}
\caption{\label{dmt_calib} $<\Delta m_t^{\rm measured}>$ and pull RMS vs $\Delta m_t^{\rm true}$.}
\end{figure}

The average of $\Delta m_t^{\rm measured}$ is compared to $\Delta m_t^{\rm true}$ and the small bias on $\Delta m_t^{\rm measured}$ is calibrated out (see Fig.\ \ref{dmt_calib}).  Similarly, we look at the RMS of the pull ($PULL = (\Delta m_t^{\rm measured}-\Delta m_t^{\rm true})/\delta( \Delta m_t )^{\rm measured}$), and the bias on the measured error on $\Delta m_t^{\rm measured}$ is removed (see Fig.\ \ref{dmt_calib}).

\subsection{Systematic Uncertainties}

Systematic uncertainties are determined by varying the composition of the pseudoexperiments.  The largest systematic errors come from signal modelling, QCD, and jet resolution.  The total systematic error of $1.15\mbox{ GeV}$ is half the size of the measured statistical error $3.44\mbox{ GeV}$, which is to say that this measurement is statistics limited.  Table \ref{tab:syst_dmt} contains a list of systematic uncertainties.

\begin{table}
\caption{\label{tab:syst_dmt}Summary of systematic uncertainties on $\Delta m_t$.}
\vspace{.25cm}
\centering
\begin{tabular}{lc}
\hline
\hline 
Source  & Uncertainty (GeV)\\
\hline
\multicolumn{2}{l}{\textit{Physics modeling}}\\
\hspace{12pt}Signal & $\pm0.85$\\
\hspace{12pt}PDF uncertainty      & $\pm0.26$\\
\hspace{12pt}Background modeling  & $\pm0.03$\\
\hspace{12pt}Heavy flavor scale factor   & $\pm0.07$\\
\hspace{12pt}$b$-fragmentation    & $\pm0.12$\\
\multicolumn{2}{l}{\textit{Detector modeling:}}\\
\hspace{12pt}$b$/light response ratio & $\pm0.04$\\
\hspace{12pt}Jet identification  & $\pm0.16$\\
\hspace{12pt}Jet resolution  & $\pm0.39$\\
\hspace{12pt}Trigger  & $\pm0.09$\\
\hspace{12pt}Overall jet energy scale & $\pm0.08$\\
\hspace{12pt}Residual jet energy scale & $\pm0.07$\\
\hspace{12pt}Muon resolution  & $\pm0.09$\\
\hspace{12pt}Wrong charge leptons & $\pm0.07$\\
\hspace{12pt}Asymmetry in $b\overline{b}$ response & $\pm0.42$\\
\multicolumn{2}{l}{\textit{Method:}}\\
\hspace{12pt}MC calibration & $\pm0.25$\\
\hspace{12pt}$b$-tagging efficiency  & $\pm0.25$\\
\hspace{12pt}Multijet contamination  & $\pm0.40$\\
\hspace{12pt}Signal fraction  & $\pm0.10$\\
\hline
Total (in quadrature)  & $\pm1.22$\\
\hline
\hline
\end{tabular}
\end{table}
 
\subsection{Measurement in Data}

The measurement in data is~\cite{Abazov:2009xq}:

\begin{eqnarray}
 e+jets & & \Delta m_t = 0.33 \pm 5.03 (stat) \mbox{ GeV} \\
 \mu+jets & & \Delta m_t = 6.74 \pm 4.71 (stat) \mbox{ GeV} \nonumber \\
 comb & & \Delta m_t = 3.75 \pm 3.44 (stat) \pm 1.15 (syst) \mbox{ GeV} \nonumber
\end{eqnarray}

\begin{figure}[h]
\includegraphics[width=40mm]{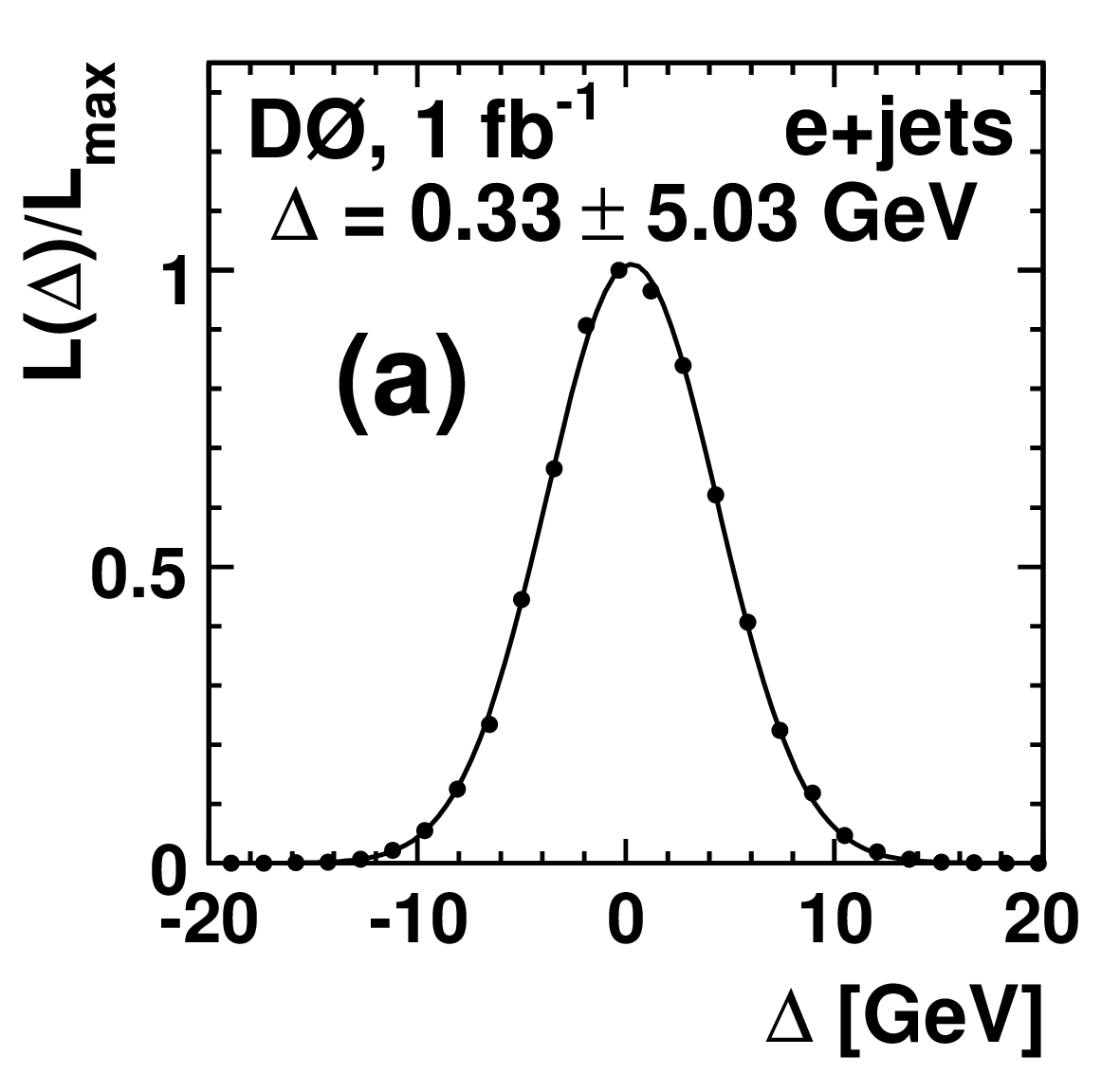}
\includegraphics[width=40mm]{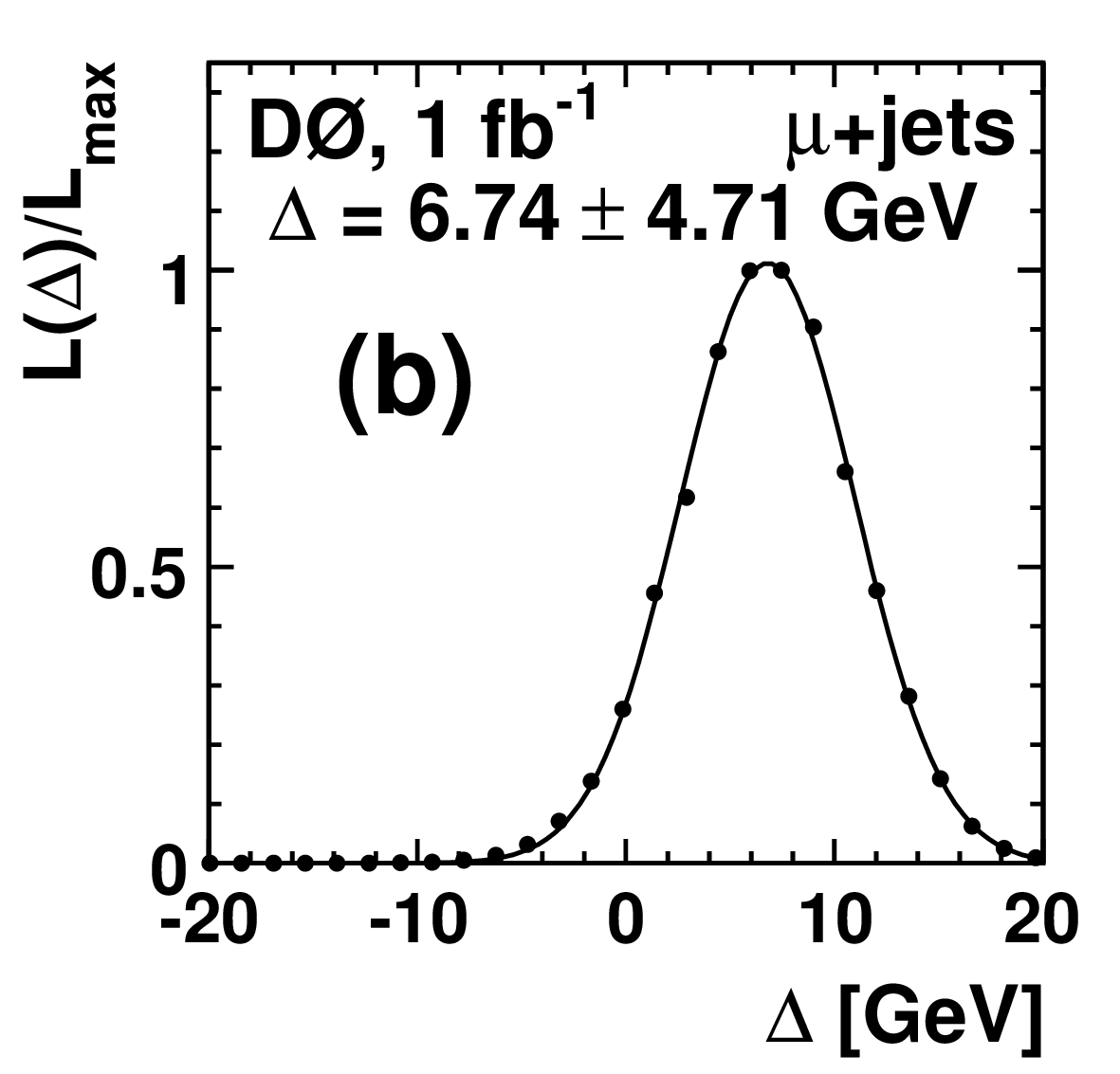}
\caption{\label{mt_data} $\Delta m_t$ measured in data for a.) e+jets b.) mu+jets.}
\end{figure}

Figure \ref{mt_data} shows the likelihood distribution in e+jets and $\mu$+jets data.

The measurement is consistent with the standard model expectation of zero, and is the first such measurement of the mass difference between a quark and its antiquark.

\section{Lepton+Jets Matrix Element Top Mass Measurement} 

The ME mass measurement uses the same method as the mass difference measurement, but with $m_t=m_{\bar{t}}$, and with the reconstructed jet energies adjusted by a constant factor $JES$, which is treated as a variable in the likelihood, and which calibrates the jet energy scale using the hadronic W mass.

\subsection{Performance on Simulated Data, Calibration of Method}

\begin{figure}[h]
\includegraphics[width=40mm]{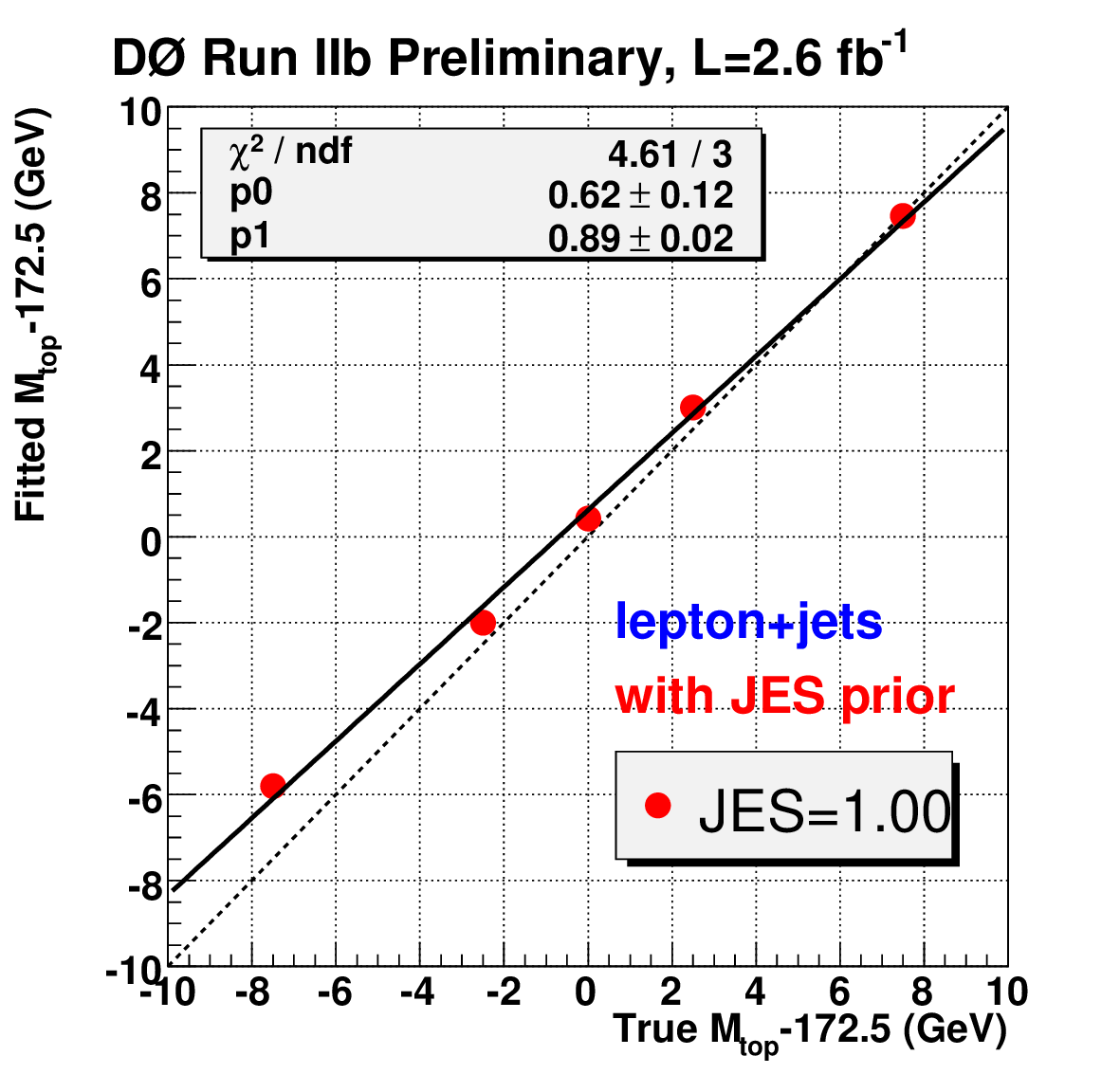}
\includegraphics[width=40mm]{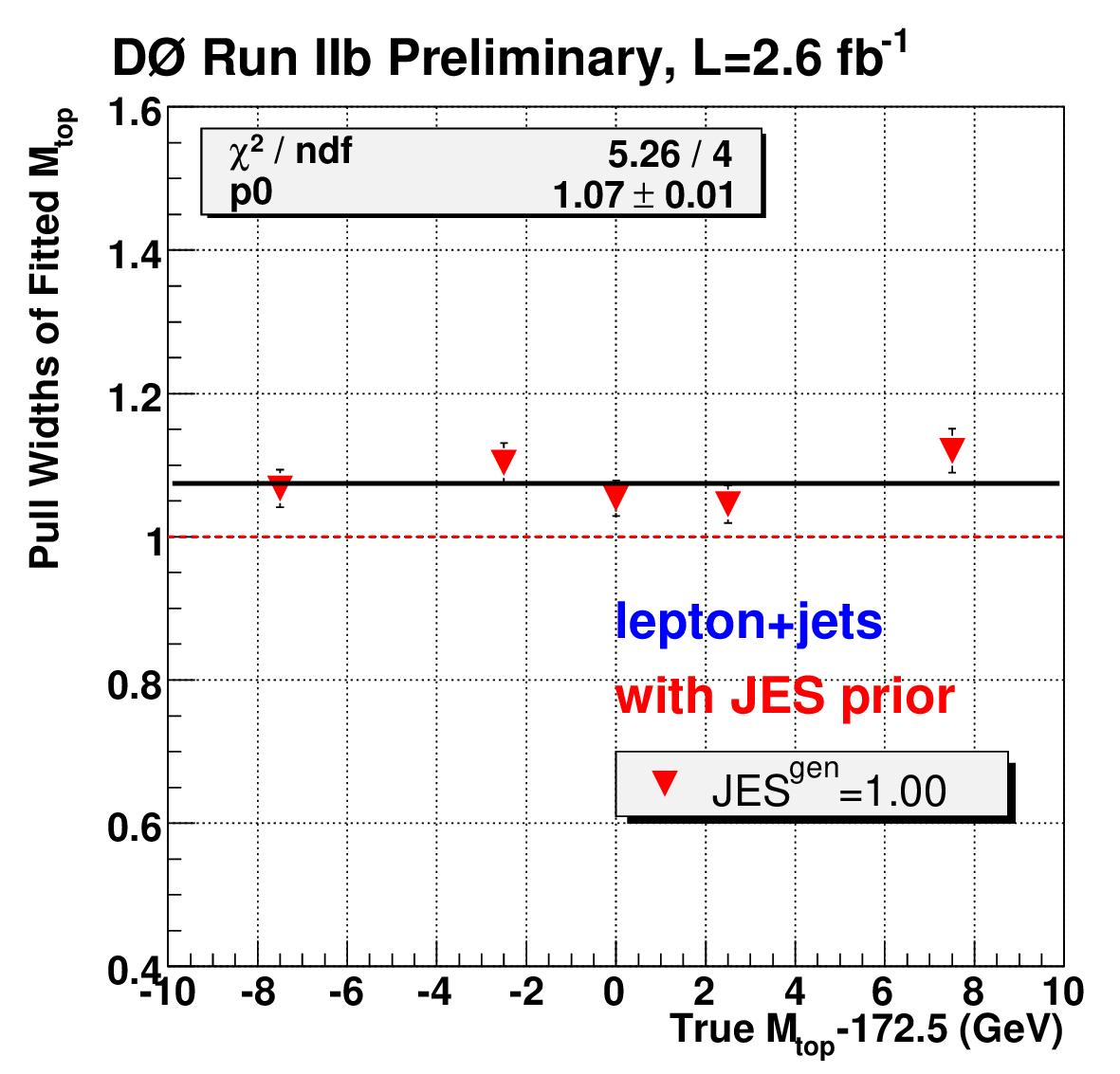}
\caption{\label{melj_calib} $<m_t^{\rm measured}>$ and pull RMS vs $m_t^{\rm true}$.}
\end{figure}

As in the mass difference measurement, ensembles of pseudoexperiments are constructed to test the performance of the mass measurement technique, and biases on $m_t$, $\delta m_t$, $JES$, and $\delta JES$ are removed (see Fig.\ \ref{melj_calib}).

\subsection{Systematic Uncertainties}

The largest systematic uncertainty come from uncertainties in the calibration of the energy of b-jets.  Table \ref{tab:syst_melj} lists the systematic uncertainties.

\begin{table}
\caption{\label{tab:syst_melj}Summary of systematic uncertainties for the lepton+jets matrix element mass measurement.}
\vspace{.25cm}
\centering
\begin{tabular}{lc}
\hline
\hline 
Source  & Uncertainty (GeV)\\
\hline
\multicolumn{2}{l}{\textit{Physics modeling:}}\\
\hspace{12pt}Signal modeling  & $\pm0.40$\\
\hspace{12pt}PDF uncertainty  & $\pm0.14$\\
\hspace{12pt}Background modeling  & $\pm0.10$\\
\hspace{12pt}$b$-fragmentation  & $\pm0.03$\\
\multicolumn{2}{l}{\textit{Detector Modeling:}}\\
\hspace{12pt}$b$/light response ratio  & $\pm0.83$\\
\hspace{12pt}Jet identification and resolution  & $\pm0.26$\\
\hspace{12pt}Trigger  & $\pm0.19$\\
\hspace{12pt}Residual jet energy scale  & $\pm0.10$\\
\hspace{12pt}Muon resolution  & $\pm0.10$\\
\multicolumn{2}{l}{\textit{Method:}}\\
\hspace{12pt}MC calibration  & $\pm0.26$\\
\hspace{12pt}$b$-tagging efficiency  & $\pm0.15$\\
\hspace{12pt}Multijet contamination  & $\pm0.14$\\
\hspace{12pt}Signal contamination  & $\pm0.13$\\
\hspace{12pt}Signal fraction  & $\pm0.09$\\
\hline
Total  & $\pm1.07$\\
\hline
\hline
\end{tabular}
\end{table}

\subsection{Measurement in Data}

The measurement in data is~\cite{Abazov:2008ds}:

\begin{eqnarray}
 \mbox{1.0 fb}^{-1} & & m_t = 171.5 \pm 1.4 (stat) \pm 1.8 (syst) \mbox{ GeV} \nonumber \\
 \mbox{2.6 fb}^{-1} & & m_t = 174.8 \pm 1.0 (stat) \pm 1.6 (syst) \mbox{ GeV} \nonumber \\
 \mbox{3.6 fb}^{-1} & & m_t = 173.8 \pm 0.8 (stat) \pm 1.6 (syst) \mbox{ GeV} \nonumber \\
\end{eqnarray}

\begin{figure}[h]
\includegraphics[width=80mm]{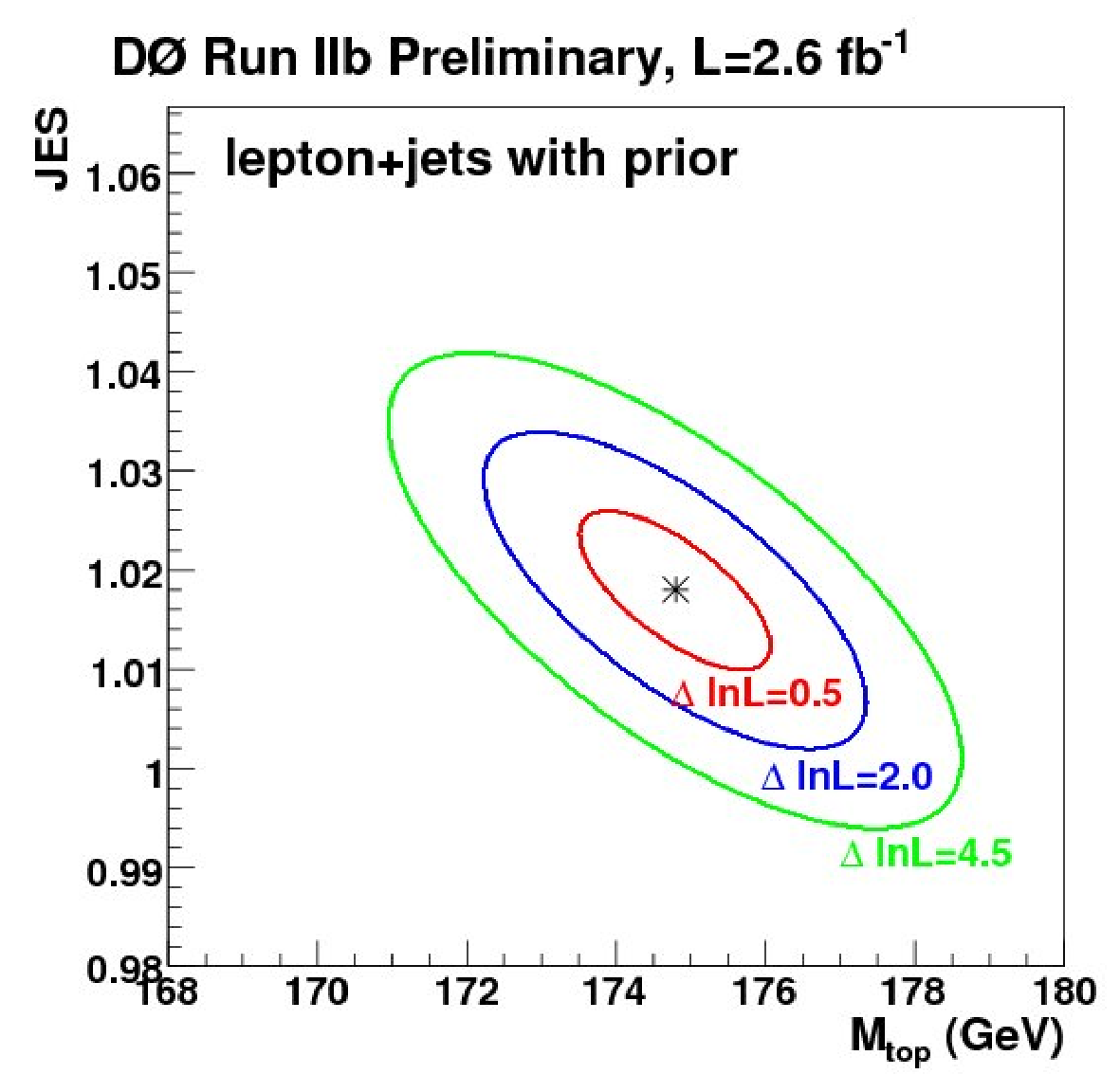}
\caption{\label{melj_data} Contour plot showing $m_t$ vs $JES$ in data	.}
\end{figure}

Here the error coming from the in-situ JES has been added to the systematic uncertainty.  Figure \ref{melj_data} shows contours of the likelihood in the $M_{top}$ vs. $JES$ plane.

\section{Dilepton Event Selection}

There are five dilepton channels, three of which use only well identified electrons and muons, the $ee$, $e\mu$, and $\mu\mu$ channels, and two which require one well identified electron or muon matched with an isolated charged track, the e+track and $\mu$+track channels.

All five channels require lepton $p_T>15\mbox{ GeV}$, all channels require two jets with $p_T>20\mbox{ GeV}$.

The main backgrounds are:
\begin{itemize}
 \item events with fake leptons (QCD, $W$+jets)
 \item $Z/\gamma^* \rightarrow ee/\mu\mu$
 \item $Z/\gamma^* \rightarrow \tau\tau \rightarrow l^-\bar{\nu}_l\nu_{\tau} l^{+'}\nu_{l'}\bar{\nu}_{\tau}$
 \item $WW\rightarrow l^-\bar{\nu}_l l^{+'}\nu_{l'}$
\end{itemize}

The background from fake leptons is controlled with lepton quality cuts, the $Z/\gamma^* \rightarrow ee$ background is controlled by requiring large $\met$ and removing events within the Z peak, the $Z/\gamma^* \rightarrow \mu\mu$ background is controlled by requiring a large missing ET significance ($\met^2/(2\sigma^2)$) where the error is calculated per event using the known resolutions of the objects in the event.  The $Z/\gamma^* \rightarrow \tau\tau$ background is reduced in the $e\mu$ channel with a cut on the sum of the leading lepton $p_T$ plus the sum of the $p_T$ of the two leading jets.

\section{Dilepton Matrix Element Mass Measurement}

The Matrix Element is employed here in the same way as for the lepton+jets measurement, the difference is that there is no JES factor applied to jets, as there is no hadronic W in dilepton events.

\subsection{Performance on Simulated Data, Calibration of Method}

\begin{figure}[h]
\includegraphics[width=40mm]{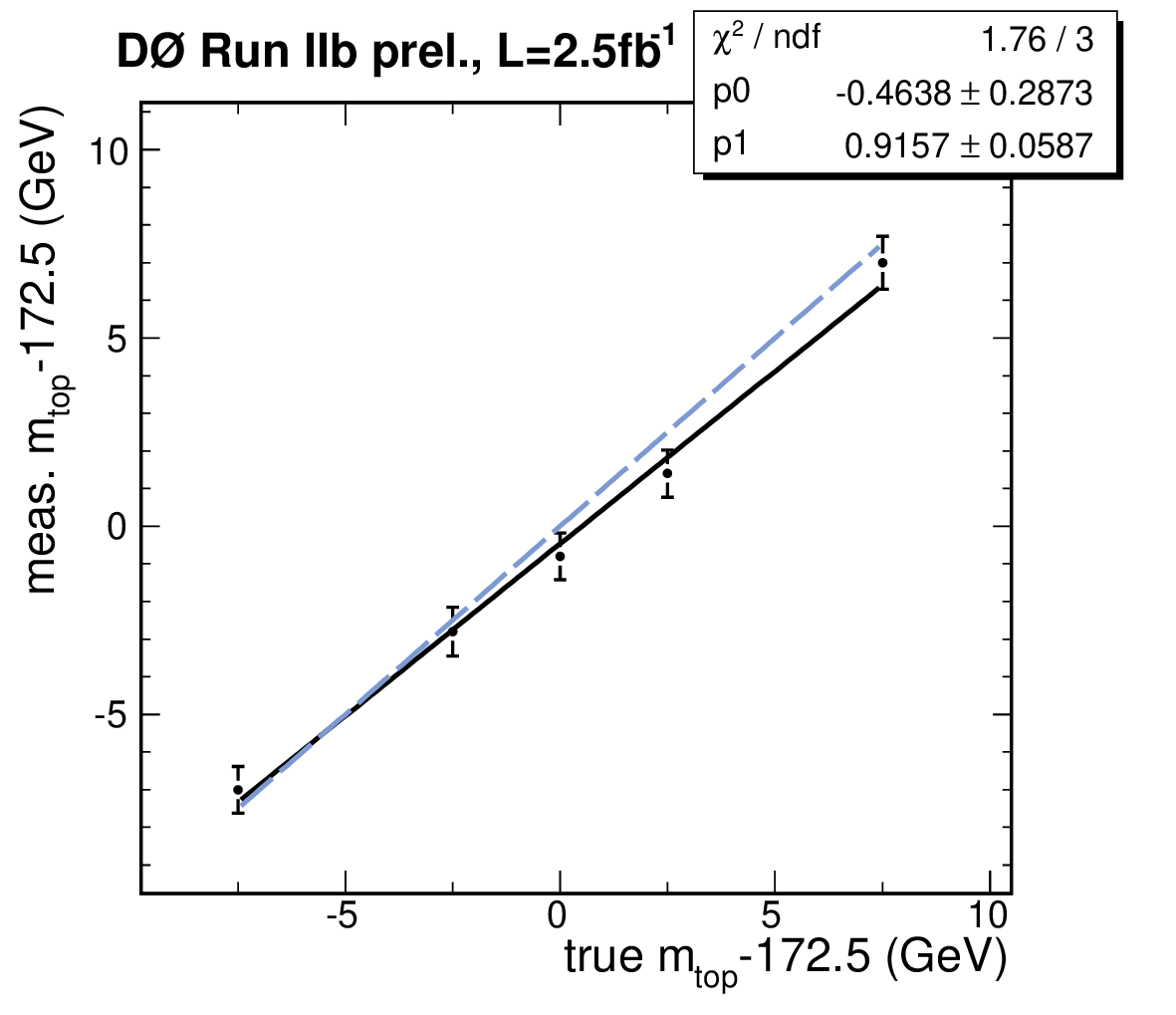}
\includegraphics[width=40mm]{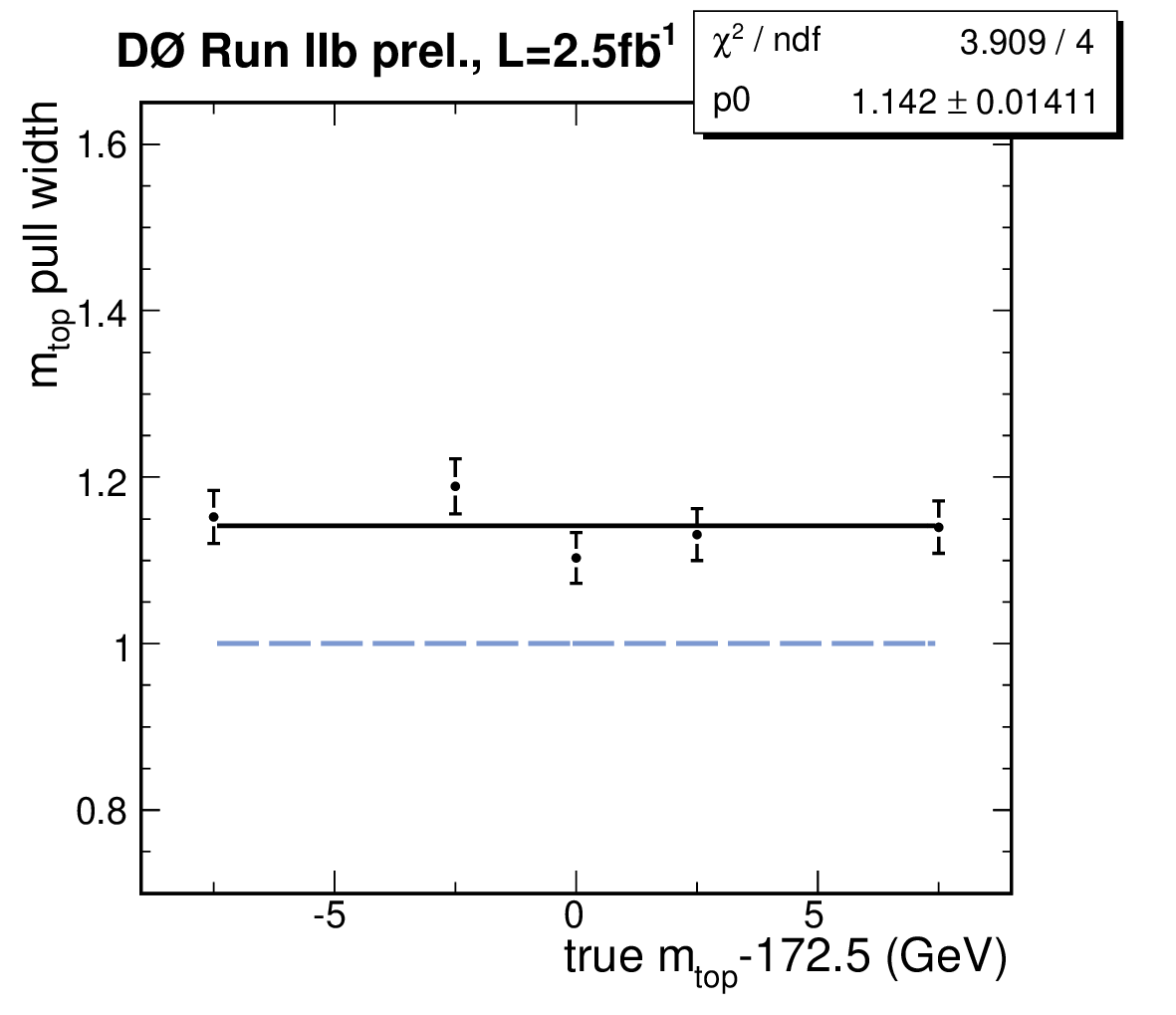}
\caption{\label{mell_calib} $<m_t^{\rm measured}>$ and pull RMS vs $m_t^{\rm true}$.}
\end{figure}

As in the l+jets mass measurement, ensembles of pseudoexperiments are constructed to test the performance of the mass measurement technique, and biases on $m_t$ and $\delta m_t$ are removed (see Fig.\ \ref{mell_calib}).

\subsection{Systematic Uncertainties}

\begin{figure}[h]
\includegraphics[width=80mm]{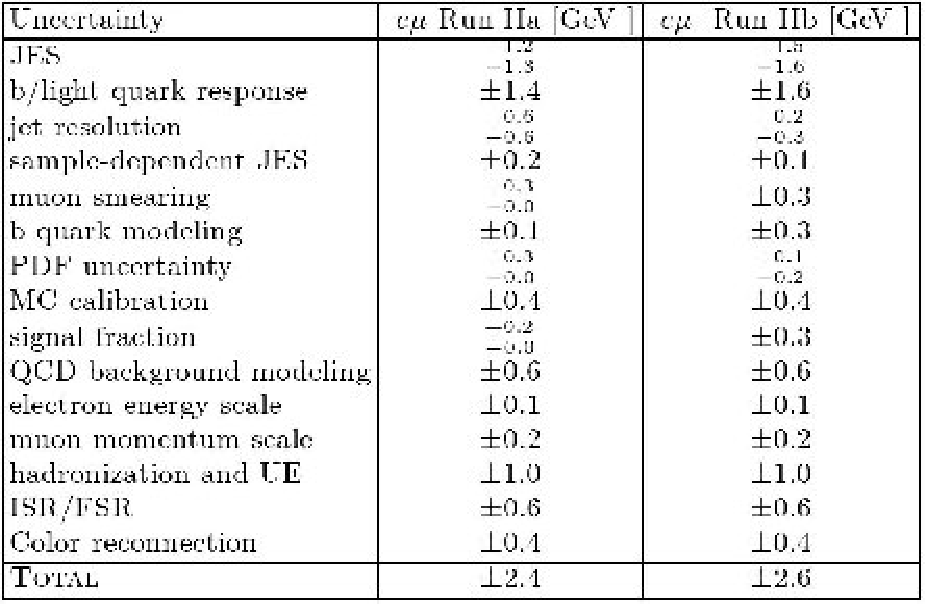}
\caption{\label{mell_syst} Systematic uncertainties for dilepton matrix element mass measurement.}
\end{figure}

The largest systematic uncertainties come from the jet energy scale calibration (both the light jet scale, and the b-jet light-jet ratio) (see Fig.\ \ref{mell_syst}).

\subsection{Measurement in Data}

The measurement in data is:

\begin{eqnarray}
 \mbox{1.0 fb}^{-1} & & m_t = 171.7 \pm 6.4 (stat) \mbox{ GeV} \nonumber \\
 \mbox{2.6 fb}^{-1} & & m_t = 176.1 \pm 3.9 (stat) \mbox{ GeV} \nonumber \\
 \mbox{3.6 fb}^{-1} & & m_t = 174.8 \pm 3.3 (stat) \pm 2.6 (syst) \mbox{ GeV} \nonumber
\end{eqnarray}

\begin{figure}[h]
\includegraphics[width=40mm]{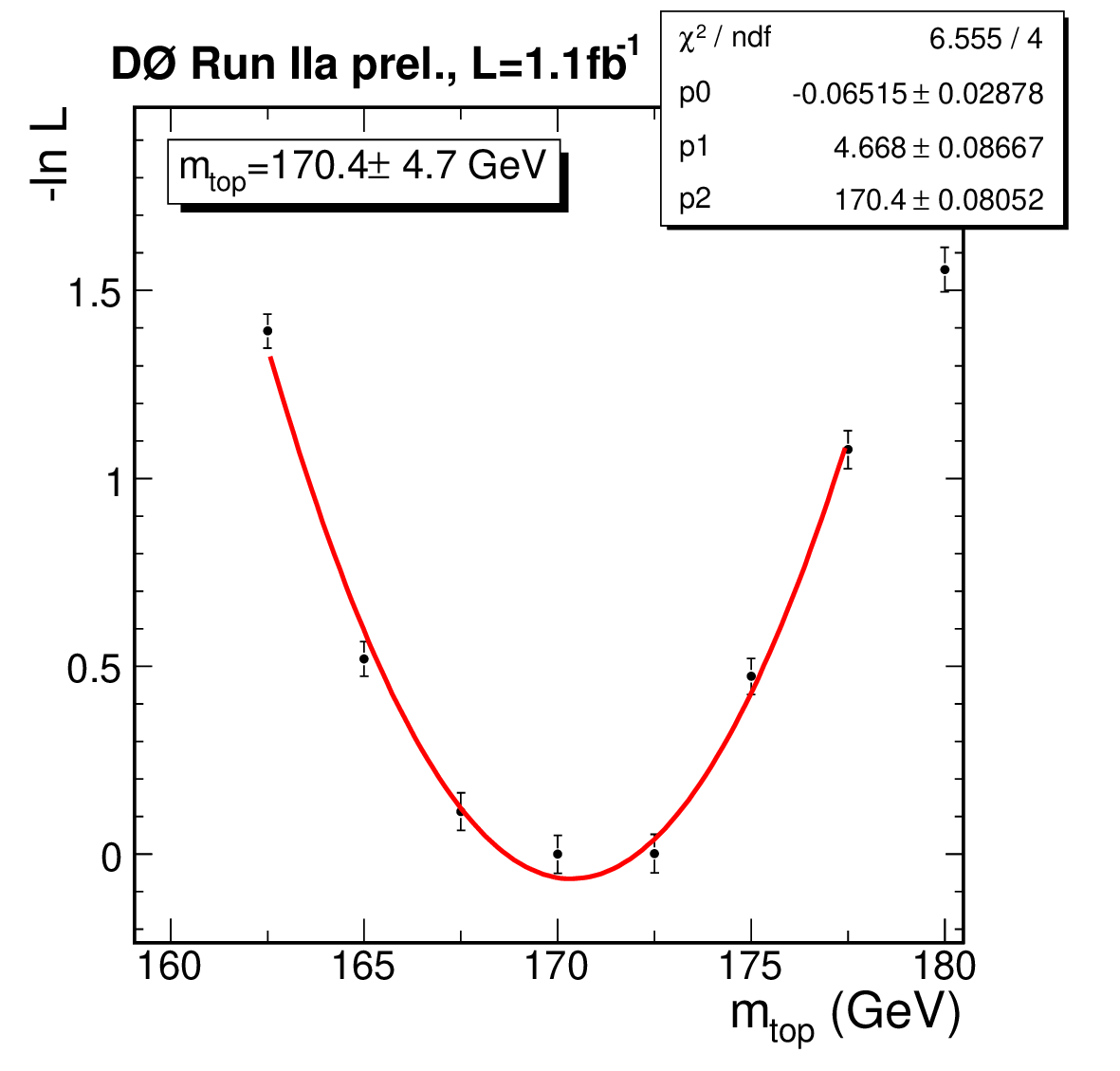}
\includegraphics[width=40mm]{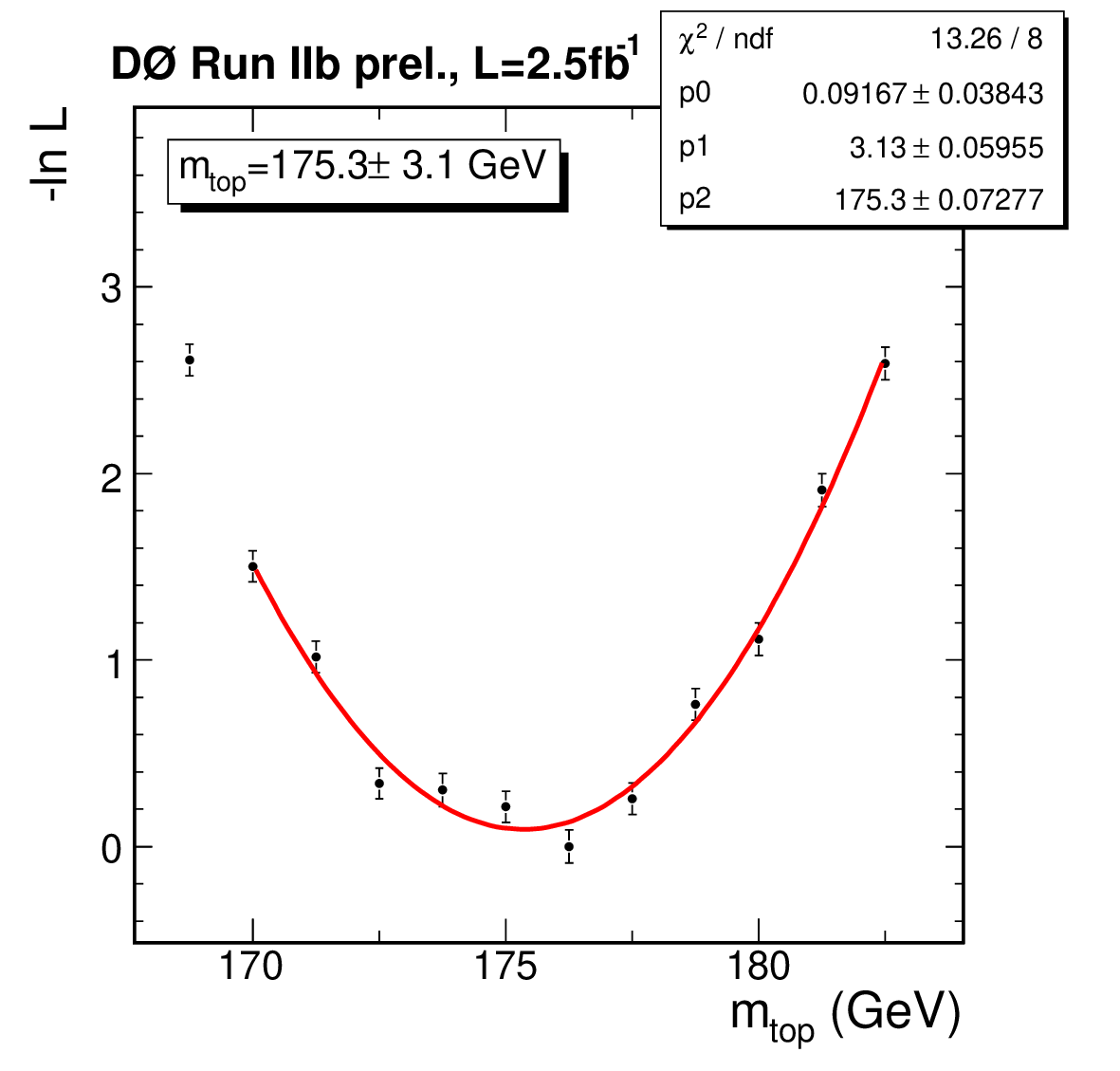}
\caption{\label{mell_data} Likelihood distribution in data for RunIIa, and RunIIb.}
\end{figure}

Figure \ref{mell_data} shows the likelihood curves for the $1\mbox{ fb}^{-1}$ and $2.6\mbox{ fb}^{-1}$.

\section{Dilepton Template-based Mass Measurements}

The two template based methods start by using the kinematic information in the event, along with a hypothesized top mass, to reconstruct one or more kinematic $t\bar{t}$ configurations consistent with the observed event.  In addition, both methods smear the lepton and jet momenta within their known detector resolutions.

The neutrino weighting ($\nu$WT) method samples the neutrino $\eta$ distribution from expected distributions, then uses this information along with the lepton and jet momenta to solve the event.  Finally, a weight is constructed which compares the $\met$ calculated using the neutrino momenta to the measured $\met$:

\begin{equation}
w = \exp\left[\frac{(\metx^{calc}-\metx^{obs})^2}{2(\sigma^{\rm UE})^2}\right] \exp\left[\frac{(\mety^{calc}-\mety^{obs})^2}{2(\sigma^{\rm UE})^2}\right]
\end{equation}

The matrix weighting method uses the measured $\met$ and uses an analytical technique to solve directly for the kinematics\cite{Sonnenschein:2005ed}\cite{Sonnenschein:2006ud}.  Each solution is then weighted by the formula:

\begin{equation}
 w=f(x)f(\overline{x})p(E_\ell^*|m_t)p(E_{\overline{\ell}}^*|m_t)
\end{equation}

Where $f(x)$ and $f(\bar{x})$ are parton distribution functions for the incoming partons, and $p(E_\ell^*|m_t)$ and $p(E_{\overline{\ell}}^*|m_t)$ give the probability for the top to decay into a lepton with energy $E_\ell^*$ in the rest frame of the top, given by the expression:

\begin{equation}
p(E_\ell^*|m_t) = \frac{4 E_\ell^* m_t (m_t^2 - m_b^2 - 2 E_\ell^* m_t)}{(m_t^2-m_b^2)^2 - m_W^2 (m_t^2+m_b^2) - 2m_W^4}
\end{equation}

\begin{figure}[h]
\includegraphics[width=80mm]{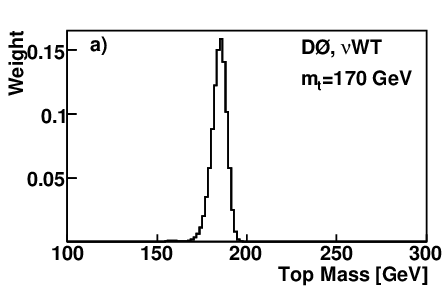}
\caption{\label{nuwt_weight} Weight distribution in $\nu$WT.}
\end{figure}

Both methods sum the weights produced at each hypothetical top mass for some number of resolution smearings (~100), and for the different solutions and jet-lepton assignments (see Fig.\ \ref{nuwt_weight}):

\begin{equation}
 W(m_t)= \sum_{N_{\rm smear}} \sum_{N_{lb}} \sum_{N_{sol}} w(m_t)
\end{equation}

\begin{figure}[h]
\includegraphics[width=80mm]{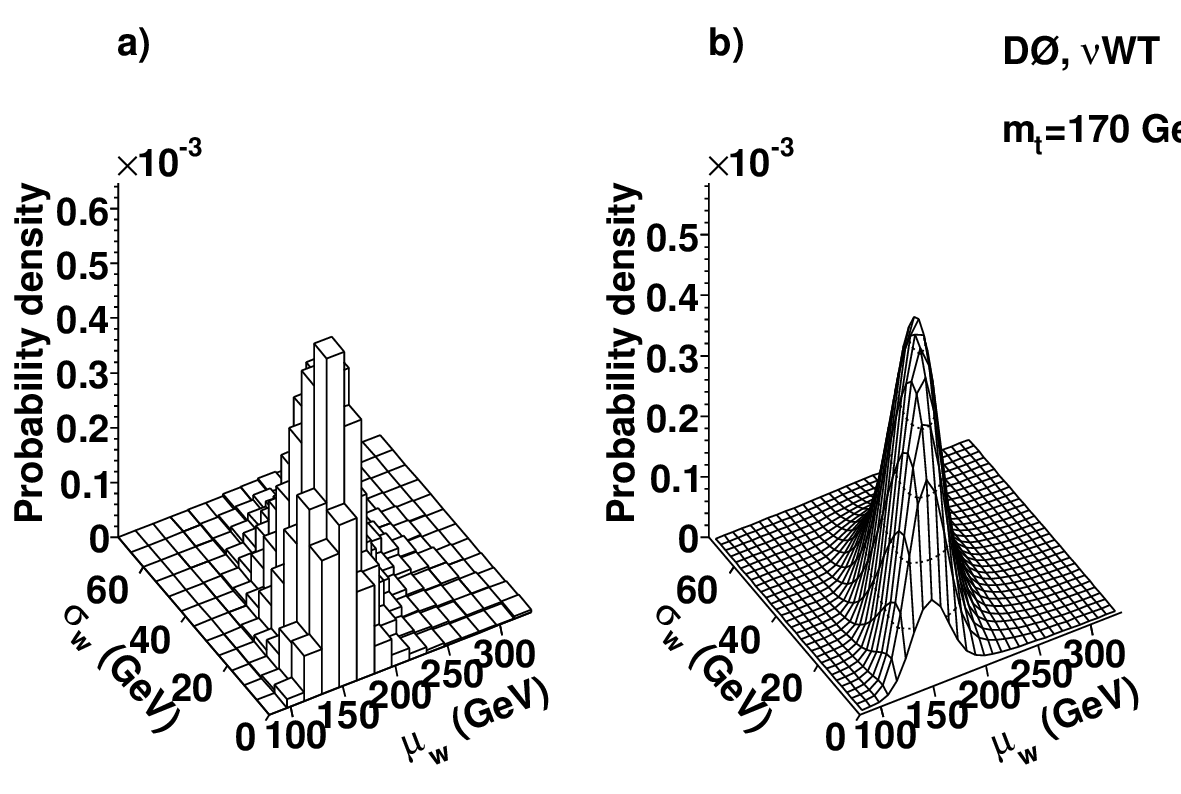}
\caption{\label{nuwt_template} Template distribution in $\nu$WT.}
\end{figure}

\begin{figure}[h]
\includegraphics[width=80mm]{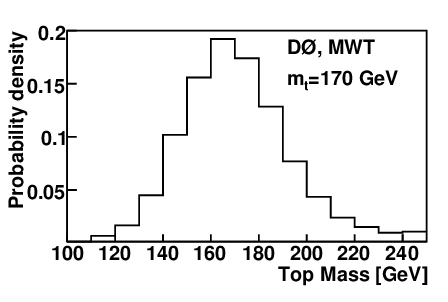}
\caption{\label{mwt_template} Template distribution in MWT.}
\end{figure}

The $\nu$WT method takes the mean and RMS of this sum-of-weights distribution, and constructs 2D distributions of signal and background known as templates (see Fig.\ \ref{nuwt_template}, note that $\nu$WT uses two correlated techniques referred to as \nuwth and \nuwtf), while the MWT method takes the mass at which the weight curve is maximized $m_t^{maxw}$, and forms 1D templates (see Fig.\ \ref{mwt_template}).  The templates are then used in a likelihood to measure the top mass.

\subsection{Performance on Simulated Data, Calibration of Method}

\begin{figure}[h]
\includegraphics[width=40mm]{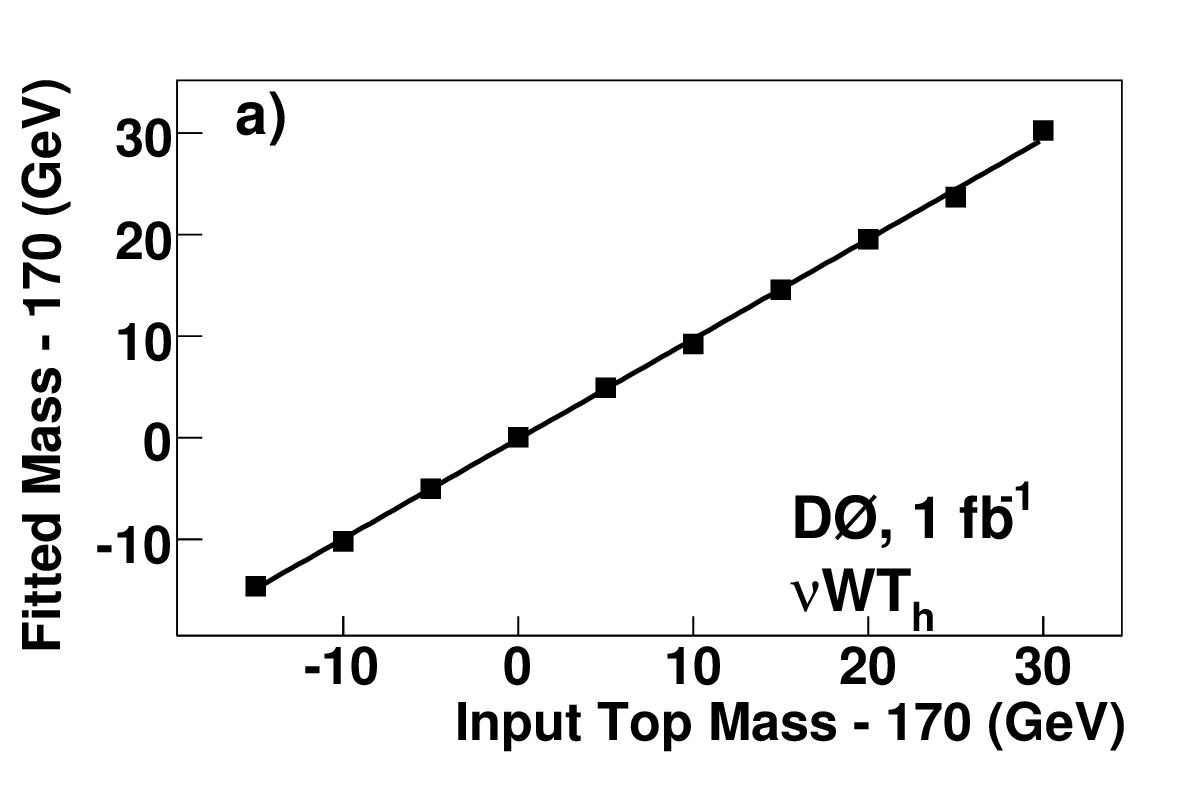}
\includegraphics[width=40mm]{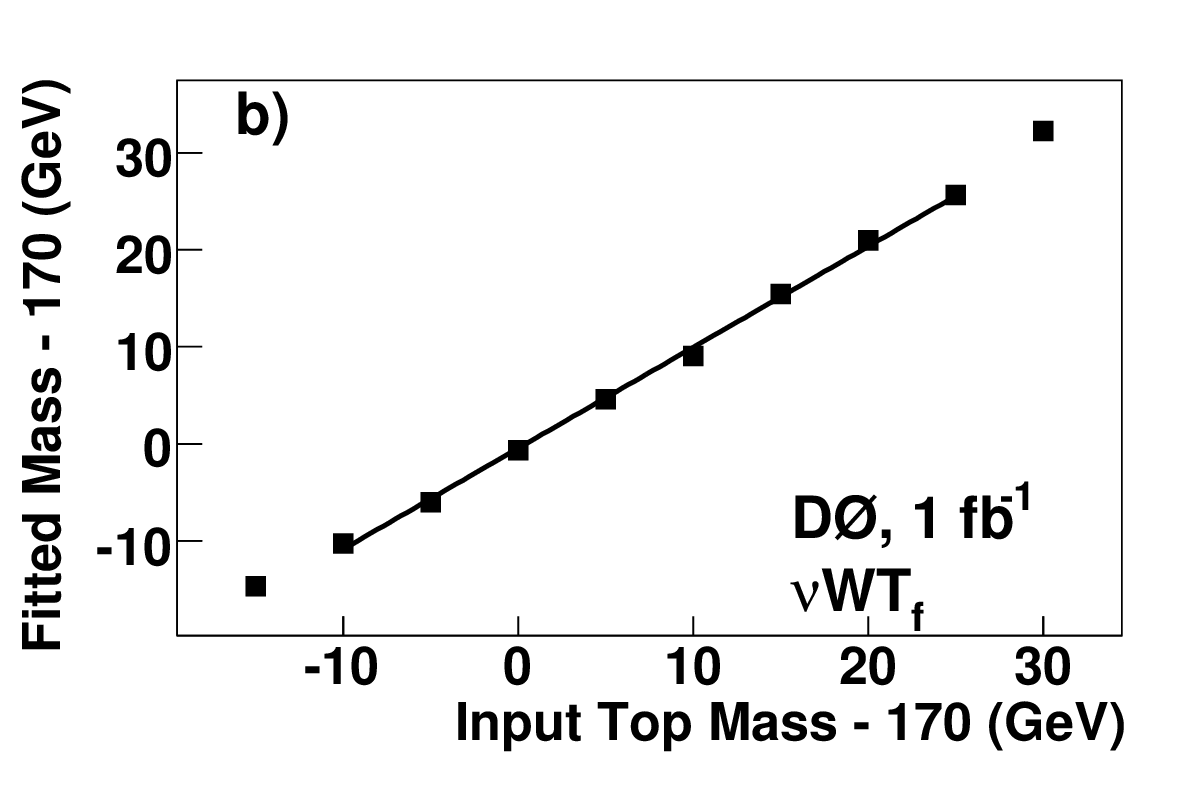}
\includegraphics[width=40mm]{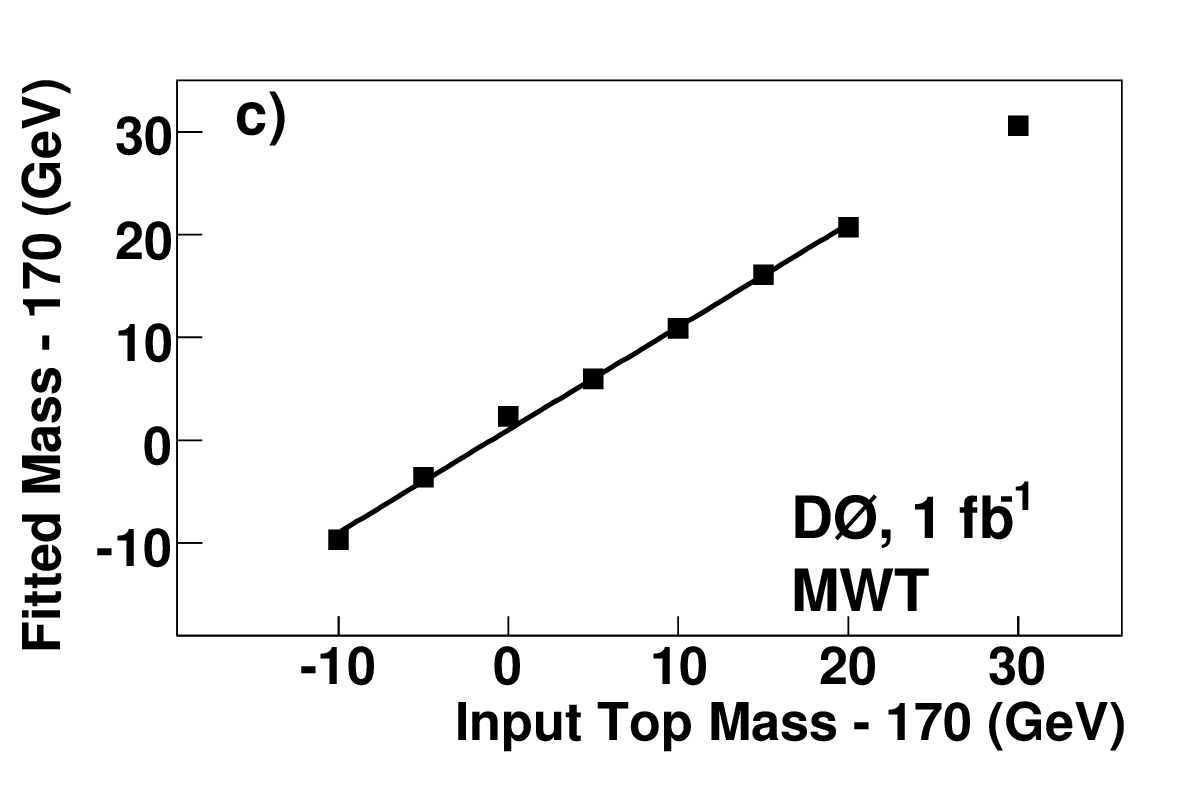}
\caption{\label{nuwt_mwt_calib} Calibration curves $\nu$WT and MWT.}
\end{figure}

As before, ensembles of pseudoexperiments are constructed to test the performance of the mass measurement technique, and biases on $m_t$ and $\delta m_t$ are removed.  Table \ref{tab:calibration} gives the slope and offset for these calibrations, while Fig.\ \ref{nuwt_mwt_calib} show the uncalibrated plots for the $\nu$WT and MWT.

\begin{table*}
\begin{center}

\caption{Slope ($\alpha$) and offset ($\beta$)
from the linear fit to the pseudoexperiment results for the $2\ell$, ltrk, and combined dilepton channel sets using the MWT and \nuwt\ methods.}
\begin{tabular}{ccccccccccccr@{.}ll}
\hline\hline
Method& Channel & \multicolumn{3}{c}{Slope: $\alpha$} & \multicolumn{3}{c}{Offset: $\beta$ [GeV]} & \multicolumn{3}{c}{Pull width} & \multicolumn{3}{c}{Expected statistical}\\
&	&		  &			  &            & & & & & & & \multicolumn{3}{c}{uncertainty [GeV]}\\
\hline
\nuwth&$2\ell$      & 0.98 &$\pm$& 0.01\phantom{0}  & $\;-$0.04 &$\pm$& 0.11 & $\quad$1.02 &$\pm$& 0.02 &$\quad\quad$ &5&8 	\\
\nuwth &\ltrk\       & 0.92 &$\pm$& 0.02\phantom{0}  & $\quad$2.28 &$\pm$& 0.27  & $\quad$1.04 &$\pm$& 0.02 & & 13&0	\\
\nuwth &combined     & 0.99 &$\pm$& 0.01\phantom{0}  & $\;-$0.04 &$\pm$& 0.11 & $\quad$1.03 &$\pm$& 0.02 & &5&1	\\
\hline
\nuwtf &$2\ell$      & 1.03 &$\pm$& 0.01\phantom{0}  & $\;-$0.32 &$\pm$& 0.15 & $\quad$1.06 &$\pm$& 0.02 & &5&8 	\\
\nuwtf &\ltrk\       & 1.07 &$\pm$& 0.03\phantom{0}  & $\;-$0.04 &$\pm$& 0.37 & $\quad$1.07 &$\pm$& 0.02 & &12&9	\\
\nuwtf &combined     & 1.04 &$\pm$& 0.01\phantom{0}  & $\;-$0.45 &$\pm$& 0.13 & $\quad$1.06 &$\pm$& 0.02 & &5&3	\\
\hline
MWT&$2\ell$         & 1.00 &$\pm$& 0.01 \phantom{0} & $\quad$0.95 &$\pm$& 0.05  &$\quad$ 0.98 &$\pm$& 0.01 & &6&3 	\\
MWT&\ltrk\          & 0.99 &$\pm$& 0.01 \phantom{0} & $\quad$0.64 &$\pm$& 0.12  &$\quad$ 1.06 &$\pm$& 0.01 & &13&8	\\
MWT&combined       & 0.99 &$\pm$& 0.01 \phantom{0} & $\quad$0.97 &$\pm$& 0.05  &$\quad$ 0.99 &$\pm$& 0.01 & &5&8	\\
\hline
\hline\hline
\end{tabular}
\label{tab:calibration}
\end{center}
\end{table*}

\subsection{Systematic Uncertainties}

As in the dilepton matrix element analysis, the dominant systematic uncertainties are the jet energy calibration and b-to-light jet ratio.  Table \ref{tab:syst_sum_combination} lists the systematic uncertainties.

\begin{table}[ht!]
\caption{\label{tab:syst_sum_combination} Summary of systematic uncertainties for the combined analysis of all dilepton channels. The \nuwth, \nuwtf, and MWT method results are shown.}
\begin{center}
\begin{tabular}{lccc}
\hline\hline
Source of uncertainty	& \nuwth\	& \nuwtf\ 	& MWT  \\
			& [GeV]		& [GeV]		& [GeV] \\
\hline
$b$ fragmentation 	& 0.4		& 0.5 		& 0.4		\\
Underlying event modeling& 0.3		& 0.1		& 0.5		\\
Extra jets modeling 	& 0.1		& 0.1		& 0.3		\\
Event generator	 	& 0.6		& 0.8		& 0.5	\\
PDF variation 		& 0.2		& 0.3		& 0.5		\\
Background template shape & 0.4		& 0.3		& 0.3		\\
Jet energy scale (JES) 	& 1.5		& 1.6 		& 1.2	\\
$b$/light response ratio& 0.3		& 0.4 		& 0.6	\\
Sample dependent JES 	& 0.4		& 0.4		& 0.1		\\
Jet energy resolution 	& 0.1		& 0.1		& 0.2	\\
Muon/track $p_T$ resolution & 0.1	& 0.1		& 0.2		\\
Electron energy resolution & 0.1	& 0.2		& 0.2	\\ 
Jet identification	& 0.4		& 0.5		& 0.5		\\
MC corrections	 	& 0.2		& 0.3		& 0.2		\\
Background yield 	& 0.0		& 0.1		& 0.1	\\
Template statistics 	& 0.8		& 1.0		& 0.8		\\
MC calibration		& 0.1		& 0.1		& 0.1		\\
\hline
Total systematic uncertainty 	& 2.1		& 2.3		& 2.0		\\
\hline\hline
\end{tabular}
\end{center} 
\end{table}

\subsection{Measurement in Data}

The measurement in data is \cite{Abazov:2009eq}:

\begin{eqnarray}
 \nu\mbox{WT} & & m_t = 176.2 \pm 4.8 (stat) \pm 2.1 (syst) \mbox{ GeV} \nonumber \\
 \mbox{MWT} & & m_t = 173.2 \pm 4.9 (stat) \pm 2.0 (syst) \mbox{ GeV} \nonumber \\
 \mbox{Combined} & & m_t = 174.7 \pm 4.4 (stat) \pm 2.0 (syst) \mbox{ GeV} \nonumber \\
\end{eqnarray}

\begin{figure}[h]
\includegraphics[width=40mm]{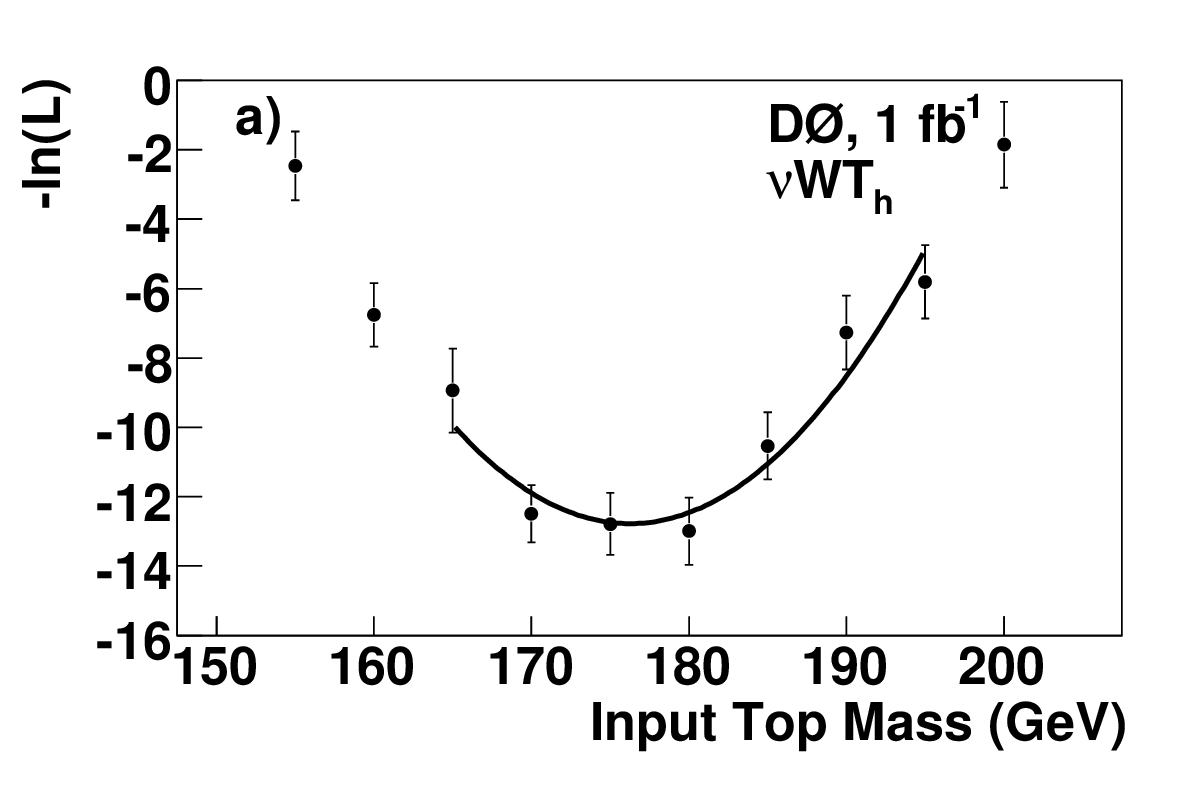}
\includegraphics[width=40mm]{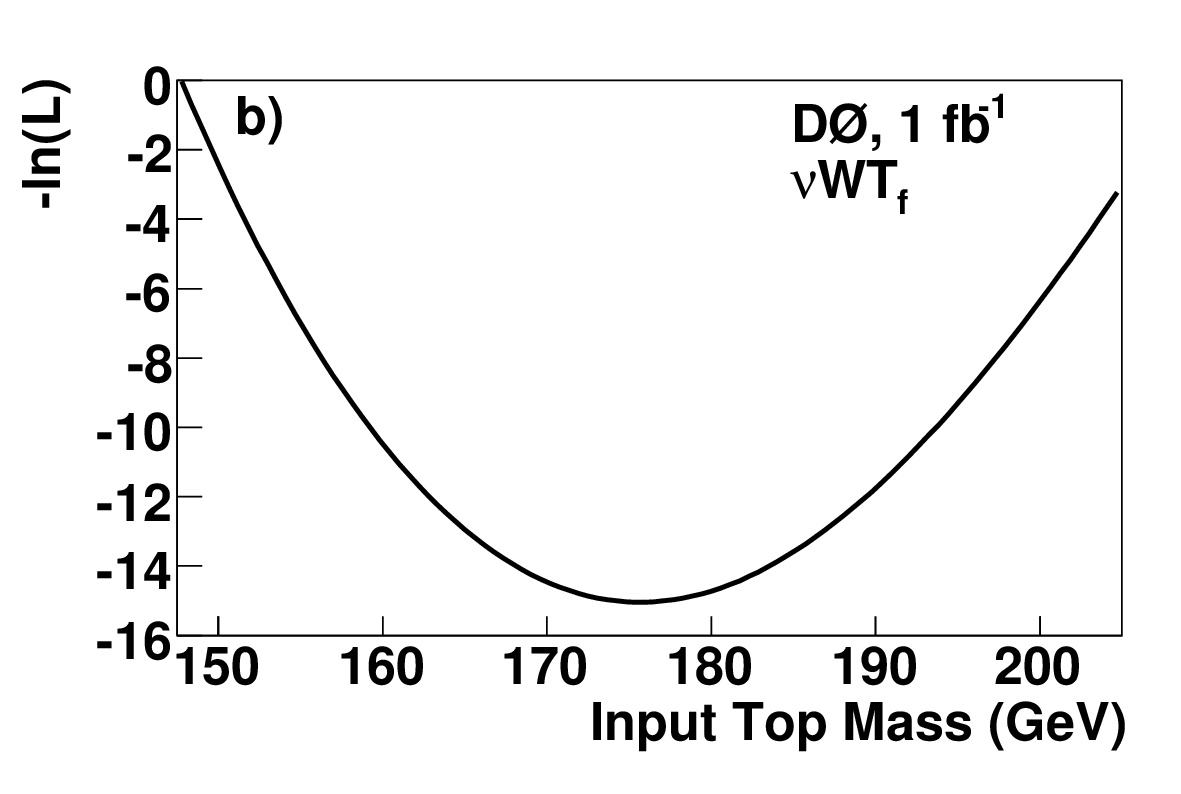}
\includegraphics[width=40mm]{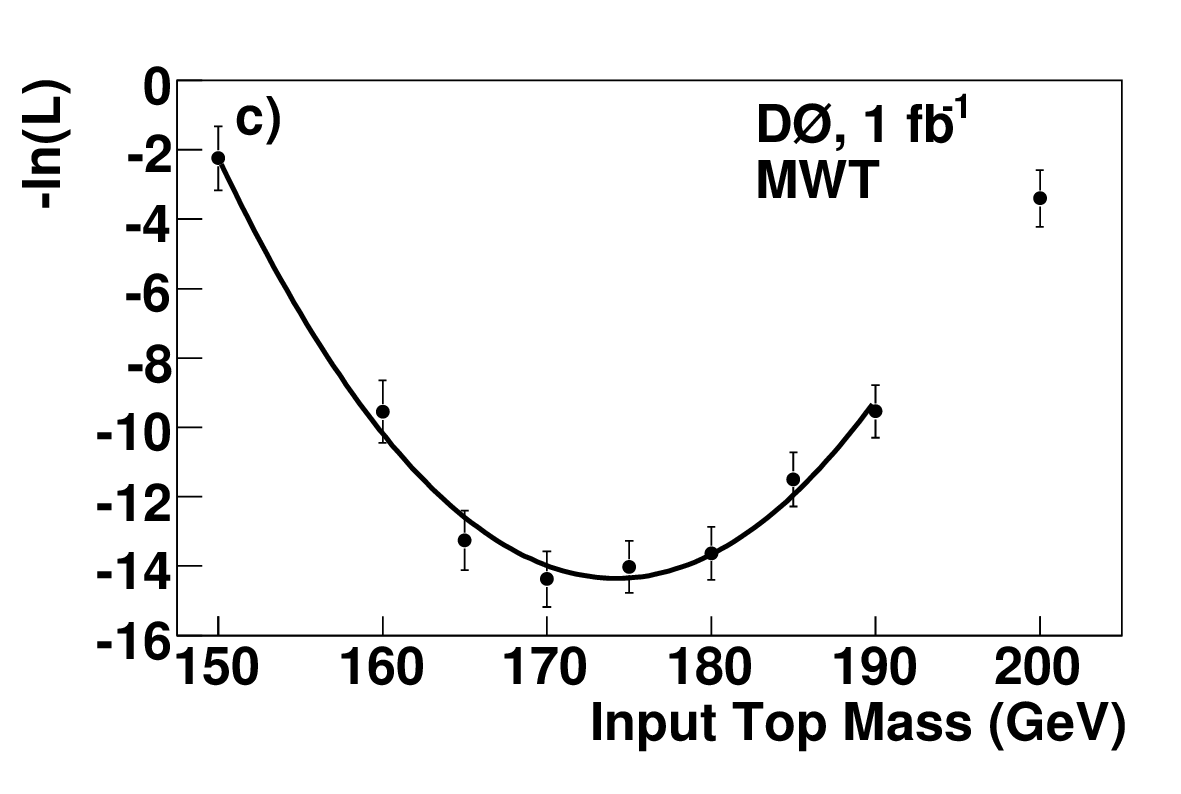}
\caption{\label{nuwt_mwt_data} Negative log likelihood distribution in data for $\nu$WT and MWT.}
\end{figure}

Figure \ref{nuwt_mwt_data} shows $-log{\mathcal{L}}$ for the $\nu$WT and MWT methods.

\section{Combination of D\O\ Top Mass Measurements, Conclusions}

\begin{figure}[h]
\includegraphics[width=80mm]{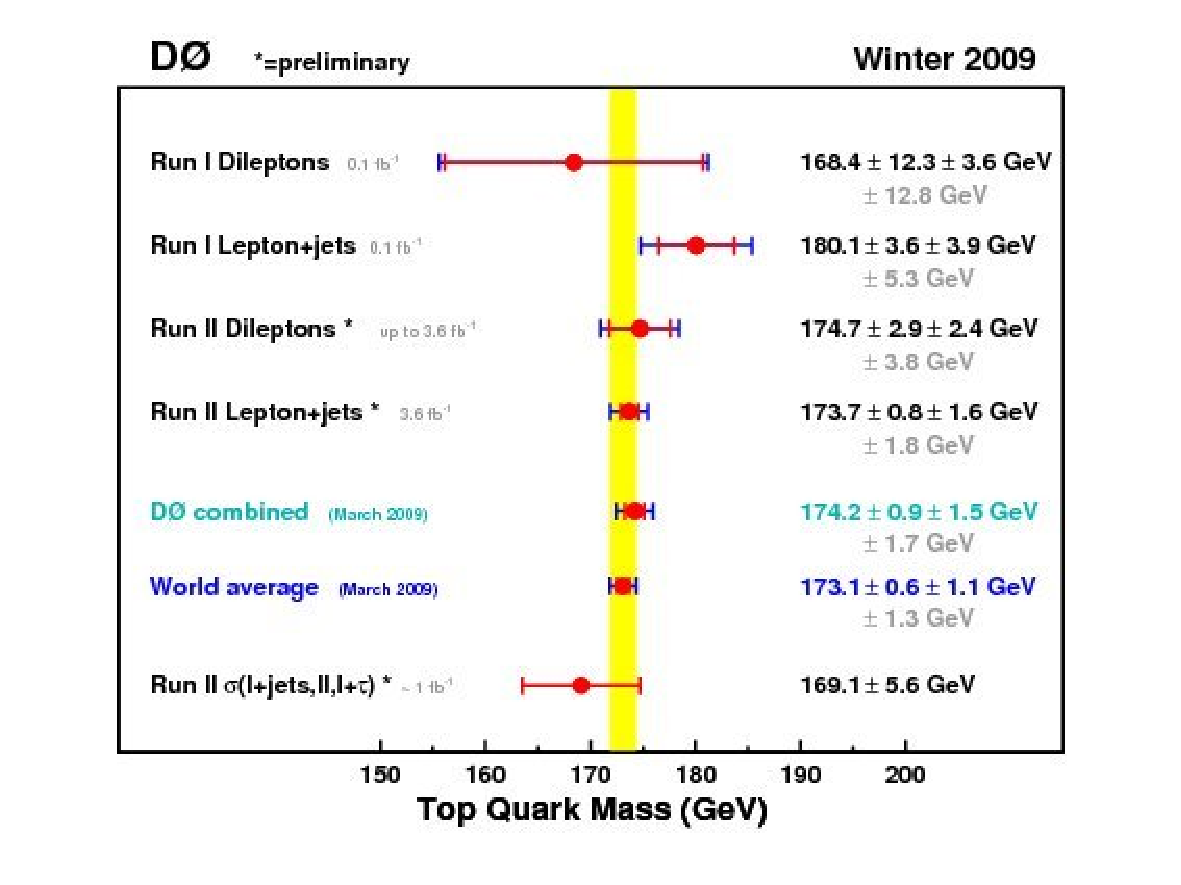}
\caption{\label{d0_mass_comb} Combination of top mass measurements at D\O\ .}
\end{figure}

The D\O\ experiment has performed the first measurement of the mass difference between the top quark and its antiquark, which is the first such measurement for any quark.  In addition the D\O\ experiment has measured the top mass to a precision of $1\%$.  Figure \ref{d0_mass_comb} summarizes the current mass measurements for D\O.

\bigskip 

\end{document}